# Two Use Cases of Machine Learning for SDN-Enabled IP/Optical Networks: Traffic Matrix Prediction and Optical Path Performance Prediction

Gagan Choudhury, David Lynch, Gaurav Thakur and Simon Tse

*Abstract*—We describe two applications of machine learning in the context of IP (Internet Protocol) /Optical networks. The first one allows agile management of resources in a core IP/Optical network by using machine learning for short-term and long-term prediction of traffic flows. It also allows joint global optimization of IP and optical layers using colorless/directionless (CD) ROADMs (Reconfigurable Optical Add Drop Multiplexers). Multilayer coordination allows for significant cost savings, flexible new services to meet dynamic capacity needs, and improved robustness by being able to proactively adapt to new traffic patterns and network conditions. The second application is important as we migrate our networks to Open ROADM networks, to allow physical routing without the need for detailed knowledge of optical parameters. We discuss a proof-of-concept study, where detailed performance data for established wavelengths in an existing ROADM network is used for machine learning to predict the optical performance of each wavelength. Both applications can be efficiently implemented by using a SDN (Software Defined Network) controller.

*Index Terms* — Machine Learning; Traffic Matrix Prediction; Multi-Layer Optimization; Routing; Open ROADMs; Optical Transport Network; SDN.

## I. INTRODUCTION

There is great recent interest in applying machine learning techniques in the networking context. See [1, 2] for recent surveys. In this paper, we provide two initial applications of machine learning to more efficiently manage IP/Optical networks in conjunction with a SDN controller.

**First Application – Predicting Network Traffic Matrix:** The traffic management of a core IP/Optical backbone of a large Internet Service Provider (ISP) must deal with dynamic traffic changes under various network conditions including scheduled and unscheduled outages, and make efficient use of network resources while also satisfying the loss and latency requirements of each class of traffic type it carries [3, 4]. It also needs to be flexible enough to provide new services demanding dynamic capacity [3, 5]. The IP layer of the network consists of IP links connected among IP devices such as router ports or white-box switch ports. The IP links are routed over a path in the optical layer using ROADMs, transponders at endpoints, and optical signal regenerators along the path when it is longer than the system optical reach. Improving efficiency means reducing the totality of IP resources (IP ports), optical resources and optical-to-electrical conversion resources (ROADMs, transponders and regenerators). If there are N traffic endpoints and K Quality of Service (QoS) classes, the totality of traffic flows can be specified by a traffic matrix of KN(N-1) elements and each such element represents the traffic from a specific source to a specific destination and belonging to a specific QoS class. The elements of the traffic matrix are usually highly correlated and their variability over time may be characterized by complex, nonlinear oscillations and seasonal periodicities at different time scales. We use machine learning for accurate short-term and long-term prediction of all elements of the traffic matrix, combine that with joint global optimization of IP and optical layers [6] using Colorless/Directionless (CD) ROADMs [7] and use Multi-Layer SDN (Software Defined Network) controller [4, 8] for implementation. This results in significant cost savings, flexible new services to meet dynamic capacity needs with better accuracy, and increased robustness by being able to proactively adapt to new traffic patterns and network conditions. The general methodology used here belongs to the category of "Traffic prediction and virtual topology (re)design" in Table II of [2]. However, while most of the related references mentioned in [2] use synthetic data, we use real data from a large ISP and furthermore combine the machine learning approach with joint global optimization of IP and optical layers.

All authors are with AT&T Labs at 200 S. Laurel Ave, Middletown, NJ 07748, USA (gchoudhury@att.com, dflynch@att.com, gt6510@att.com, stse@att.com)

**Second Application – Predicting Optical Path Performance in a Multi-Vendor Network:** Large ISPs typically operate single-vendor Layer 0 ROADM networks for optical transport. Before provisioning new wavelengths, the ISP should verify that the proposed physical routes meet optical performance standards. In this paper we use the term "wavelength" to mean "wavelength connection". Usually, this evaluation is conducted using closed vendor-proprietary tools that incorporate detailed analysis of the various vendor specific optical components. An Open ROADM network architecture initiative has been launched [9-11] where the ROADMs and other optical plug-ins will be model-driven with open standard interfaces, thus allowing interoperability among different vendor equipment. The introduction of Open ROADM and the SDN controller technologies will allow ISPs to more effectively and uniformly leverage network performance data to set up optimal wavelength paths that meet optical performance standards. Because Open ROADM will integrate equipment from multiple vendors, single-vendor performance evaluation tools will no longer be suitable for evaluating new wavelength paths. Instead, we propose a new machine learning model that will use network data to predict optical performance of new wavelengths in a multi-vendor environment. We describe a proof-of-concept study, where we collect detailed information for established wavelengths in an existing ROADM network, and then use machine learning to predict the optical performance of each wavelength, specifically the bit error rate. The machine learning model is able to predict the bit error rate with an acceptably small mean squared error value. It can be incorporated into the Path Compute Engine (PCE) within the SDN Controller to verify that all new Open ROADM wavelengths meet optical performance standards. The model can also monitor the performance of existing wavelengths and proactively move and/or groom them to better paths as conditions evolve. The general methodology used here belongs to the category of "QoT (Quality of Transmission Estimation)" in Table I of [2] and is perhaps closest to [12]. However, while [12] (and most of the related references in [2]) uses synthetic data, we use real data from a large ISP's network. Also while the model in [12] predicts whether the optical performance of a new wavelength will be good or bad, we predict actual BER allowing use of different thresholds depending on what the new wavelength will be used for and allowing comparison of alternate wavelength paths.

## II. Machine Learning For Traffic Matrix Prediction In a Core IP/Optical Network

### A. Framework for Closed Loop Optimization using Machine Learning

Figure 1 depicts the framework for self-optimizing an IP/Optical network in a closed loop manner where future traffic prediction from machine learning, real-time network and traffic measurements, and knowledge based feedback on traffic changes and failures will collectively drive a joint global optimization engine for both the packet and optical layers. A multi-layer SDN controller collects long-term and short-term traffic and failure data to facilitate these three types of feedback, implements the global optimization algorithms, and pushes the required changes to the packet and/or optical layers of the network. Optimization needs to be done on at least two different time scales. In the short time scale (seconds, minutes and hours) the available network resources are fixed and we have to use them optimally. In this setting, network changes (traffic matrix and network failures) can be detected and the network is re-configured (either at the packet or at the optical level) in a reactive mode based on real-time feedback. Short-term traffic prediction based on machine learning allows us to respond to these changes in a more resource-efficient and less disruptive way. In the longer term (days, weeks and months), we need to perform a network design exercise including simulation of many potential traffic change and failure scenarios to determine the optimal level of resources. Here long-term traffic prediction based on machine learning will play a key role.

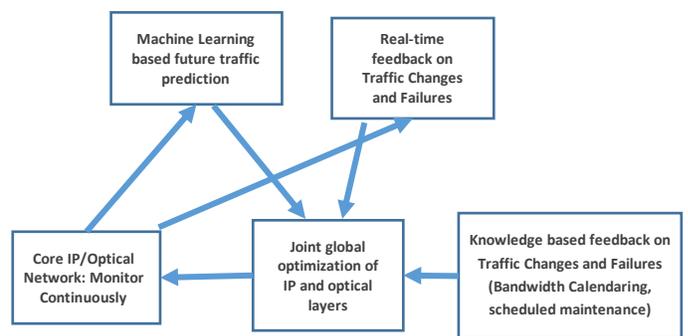

Fig. 1. Closed Loop Optimization with Machine Learning and other Feedbacks.

### B. Routing of traffic over Packet/Optical Network

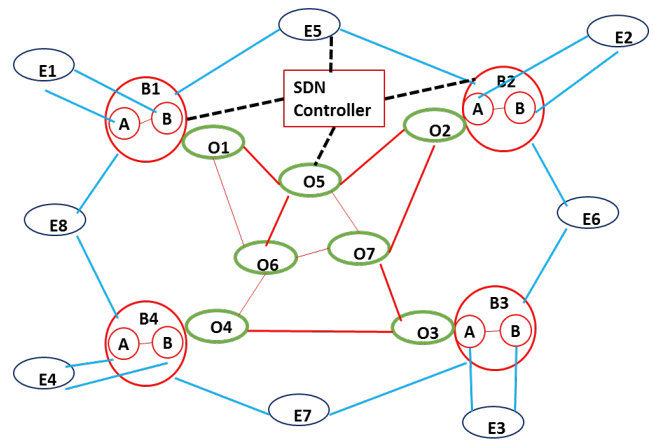

Fig. 2. Example IP/Optical Network with Backbone (B), Edge (E) and optical (O) locations.

Figure 2 illustrates an example of an integrated IP/Optical network and its interaction with the SDN controller. $E_i$ represents the IP edge routers, $B_i$ represents IP core or backbone locations and $O_i$ represents optical nodes (ROADMs). A subset of the optical nodes is collocated with an IP core location. We show two core routers, A and B, per core location but in general the number can be variable. All unicast traffic originates/terminates at the edge routers and each such router is connected to at least two core routers (same or different locations) using physically diverse paths.

A subset of all possible pairs of IP routers is connected via IP links to form an IP network. An IP link between two different core locations needs to be routed over the optical network. As an example, the IP router A in location B2 may be connected to the IP Router B in location B4 over the sequence of optical nodes O2-O7-O6-O4. The SDN controller can control the edge routers, the core routers and the optical nodes. The SDN controller is logically shown as a single centralized entity but it may be functionally separated into one controlling the IP network and one controlling the optical network. Furthermore, for the purpose of reliability and disaster recovery, it makes sense to have one active SDN controller and one or more standby SDN controllers located geographically in different places.

Figure 3 explains various levels of routing in the network. The IP links are routed over the ROADM layer and the MPLS TE tunnels carrying end-to-end traffic are routed over the IP layer.

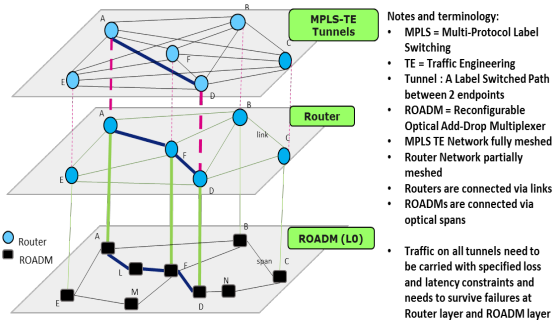

Fig. 3. Various Levels of Routing.

## C. Flexibility of Resource Management with CD ROADMs and Digital Fiber Cross Connect (DFCC) devices

Figure 4 shows an end-to-end routing of an IP link over the optical layer. R1 and R2 represent router ports in two different geographical locations. T1 and T2 are transponders used in the two locations for electrical-to-optical and optical-to-electrical signal conversions. The connected combination of a router port and transponder in the same location is called a Tail and we have two tails in this illustration: Tail1 and Tail2. There may be many ROADMs in the optical network (ROADM1 is collocated with Tail 1 and ROADM4 is collocated with Tail 2). In addition there may be one or more electronic regenerators (e.g., RE1 as shown in the picture) needed to boost signal strength if the route-miles from ROADM1 to ROADM4 is beyond the system optical reach. We may also use a Dynamic Fiber Cross-connect (DFCC) device [13, 14] to connect the two components of the Tail in the same location. There are usually many router ports and transponders in the same location (not shown in the picture) and the DFCC allows us to connect any router port to any transponder and dynamically rearrange the connections following a traffic change or failure event.

Traditionally, if any component along the path of the IP link fails (or there is a fiber-cut), the entire IP link fails and no non-failed component can be reused. However, with an SDN controller managing both the packet and the CD ROADM networks [8], the three components, namely Tail1, Tail2 and RE1 are disaggregated, interoperable and the non-failed components can be reused by the controller. Furthermore, the DFCC device also disaggregates the two components of the Tail and if one of its components fails, the non-failed component can be reused and combined with another component of the opposite type to form a new Tail. The real-time SDN controller can leverage this resource disaggregation capability to provide numerous resource reuse/sharing opportunities to proactively overcome traffic fluctuations and network failures.

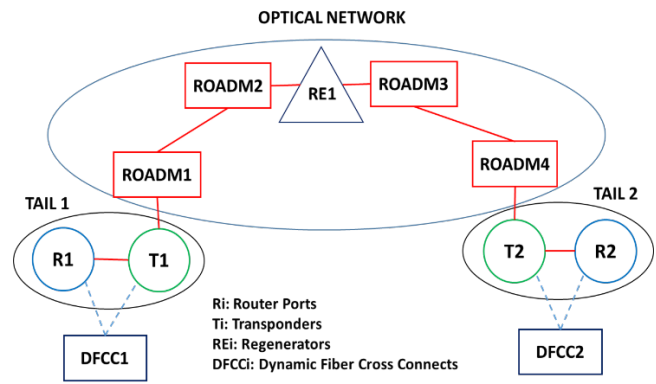

Fig. 4. End-to-end Routing of IP Link over Optical Layer.

## D. Machine Learning-Based Future Traffic Prediction

If there are N traffic endpoints and K QoS classes then there are $T = KN(N-1)$ elements in the traffic matrix. As an example, if $K = 2$ and $N = 50$ then $T = 4900$. We assume that each element of the traffic matrix is routed over the packet network as a TE (Traffic Engineering) tunnel. In general, for routing flexibility, each TE tunnel may be split into multiple ones but we ignore this here for simplicity and illustration purpose. Typically, the TE tunnel traffic at a large ISP network is characterized by complex, nonlinear oscillations and seasonal periodicities at different time scales, reflecting customer usage of the network. The traffic on the highest-activity tunnels contains a strong daily oscillation, a less prominent weekly oscillation (reflecting different usage patterns on weekends), along with occasional sharp jumps that correspond to the network dynamically shifting traffic between tunnels following an IP topology change (the sharp jumps are not directly predictable by the forecast model, but the model adjusts to the new level of the data immediately on the next time point after observing the jump. Furthermore, since we usually know about the long-term IP topology changes and associated routing changes ahead of time, we can feed that information to the prediction model and improve accuracy). An example of the total traffic volume and the traffic volume on a particular TE tunnel is shown in generic bandwidth units in Figure 5. Our goal is to develop a machine learning based, real-time prediction of the traffic load for each of the TE tunnels at future time horizons of minutes, hours, days and weeks, although we primarily concentrate on hours or longer time scales.

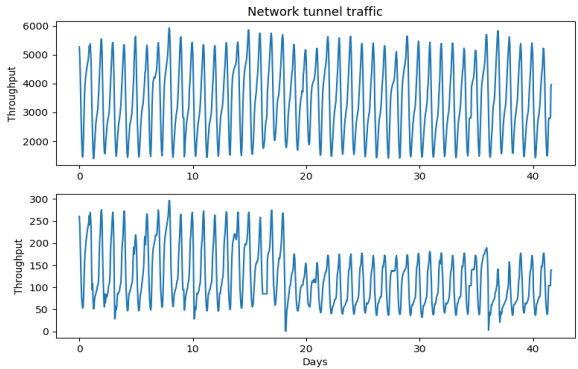

Fig. 5. The tunnel traffic volume across the entire network (top) and an individual TE tunnel (bottom) using generic bandwidth units.

We denote a given TE tunnel's traffic by $\{x_0, x_1, \ldots, x_N\}$. For each TE tunnel, the goal is to form a statistical model that forecasts the traffic on that TE tunnel for a given forward time horizon $a_T$, such as the next hour or next 24 hours. We use a nonlinear autoregressive-like model of the form $x_t = f\left(x_{t-a_1}, x_{t-a_2}, \ldots, x_{t-a_T}\right) + b + ct + \epsilon_t$, where $\{a_1, \ldots a_T\}$ are pre-specified time lags, $b + ct$ is a linear trend and $\epsilon_t$ is Gaussian white noise. We estimate the mapping $f$ by applying Gaussian process regression (GPR), a Bayesian nonlinear regression model where the number of parameters estimated grows with the amount of data. GPR, also known as Kriging, models $f$ as a realization of a Gaussian process with a covariance kernel function formed from the observed data. The posterior estimate $E(x_t|f)$ under the model has an explicit formula in terms of the training data, and can be used to make out-of-sample predictions. GPR is used extensively in different fields and has been found to perform well in situations with limited data available, although the standard form of the algorithm can become computationally intensive as the data size grows. GPR can also be viewed as a probabilistic formulation of kernel regression and provides Bayesian credible intervals (error bars) on any forecasts, which are helpful in interpreting the results. In comparison to a classical, linear autoregressive (AR) or moving average (MA) model, this type of model is better able to capture the asymmetry between the rising and falling parts of the daily oscillation, as seen in Figure 5. More details of GPR can be found in [15].

We apply GPR by first de-trending the time series with a linear regression, and then regressing $\{x_{t-a_1}, x_{t-a_2}, \ldots, x_{t-a_T}\}$ on $x_t$ for all $t$ such that both $t$ and $t - a_T$ lie within the training period. In the machine learning literature, GPR is typically applied to time series in a different manner than this, by regressing $t$ on $x_t$ for all times $t$ in the training dataset. However, this approach requires more detailed prior knowledge of the data to specify a good kernel function (a key part of the GPR model), and also lacks a direction of time or notion of causality in the model. In practice, we found that it performed worse than the lagged approach described above. We also applied several other regression models that are standard in the machine learning literature, including penalized linear models, boosted decision trees and random forests (see [16] for details), but GPR was found to have better out-of-sample prediction accuracy than these other methods. This is likely explained by the fact that our training data size is limited but has a relatively high signal-to-noise ratio and the mapping $f$ is stationary over time, which is well suited for GPR.

In practice, the choice of lags $\{a_1, \ldots a_M\}$ has a large impact on the model's accuracy. Specifying too few lags fails to capture longer term dependencies in the model, while having too many lags results in a large parameter space where $f$ cannot be estimated efficiently. It is known that GPR generally becomes less accurate for data with a large number of features. To choose the lags, we apply a heuristic based on the partial autocorrelation $\rho_t$ of the data. The partial autocorrelation has been widely studied in classical time series analysis and is used in the Box-Jenkins methodology [17] to find the number of lags in a classical AR model. It is defined by $\rho_t = corr(x_0, x_t \,|\{x_1, x_2, \ldots, x_{t-1}\})$, and represents the amount of extra correlation at time lag $t$ after accounting for the correlation at all smaller lags. An example of $\rho_t$ for the total tunnel data (top half of Figure 5) is shown in Figure 6. We compute $\rho_t$ over the training period, and choose only those lags where $\rho_t > a_T^{1/6}/15$. The intuition behind this choice is that for small $a_T$ (say, one hour ahead), the mapping $f$ is easy to estimate and we can estimate a higher dimensional model with more lags, while for large $a_T$ (one week ahead), the mapping $f$ is much noisier, and we estimate a lower dimensional model to compensate for it.

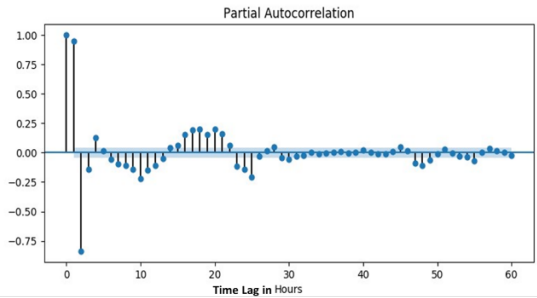

Fig. 6. Partial autocorrelation of total network traffic over three months of data with the time lag ranging from 0 to 60 hours. The shaded region is a 95% confidence interval.

We use data collected from a large ISP network over 2017, with the data up to 11:00 PM, July 31 used for training the model, and the data from (12:00 AM, August 1) + ($a_T$ hours) onward used for testing the model's performance, where $a_T$ is the desired forecast horizon. The gap between the training and testing periods ensures that there is no overlap in the data used to train the model and the data used to test it. This is done for each tunnel separately, using its own past history, and for a range of different $a_T$ from 1 hour to 168 hours (one week). The choice of lags is determined separately for each tunnel and value of $a_T$. The tunnels with the highest activity have quite different characteristics than the ones with lower activity, and typically result in a different choice of lags. We use the scikit-learn implementation of GPR [18] with a squared-exponential kernel in GPR (a standard choice; see [15]), with a bandwidth parameter $\theta = 0.01$ and a noise power $var(\epsilon_t) = 0.01$. For a given model, we measured the error using the relative median absolute error (MAE) on the test period of the data, as well as the relative MAE over only the period of peak activity in the network, 1:00 AM to 5:00 AM GMT, which is important for capacity planning purposes. The MAE is a more appropriate metric than the standard mean-squared error since it is less sensitive to error contributions from short impulses or bad data points. We train the models for each TE tunnel from May to July and test them over August. For the total traffic, the relative MAE over the whole test period is 1.61% and the relative MAE over the peak periods is 1.12%. For individual TE tunnels, the error metrics are 2-10% for the high activity ones and 5-30% for lower activity ones (where the traffic often consists of random impulses that are not predictable). As a point of comparison, a linear AR model was typically found to achieve an MAE of 6-15% on the high activity tunnels and 9-30% on the low activity ones.

An example of the forecasted total traffic for several different $a_T$ is shown in Figure 7, with the different models combined to form a forecasted trajectory over four days (essentially 96 different models, each one trained and used for a different forecast horizon between 1 and 96 hours).

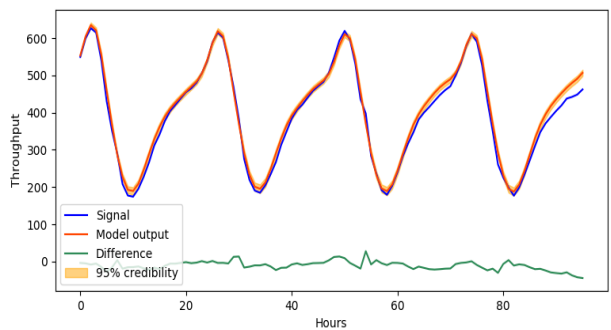

Fig. 7. Comparison of the total traffic (blue) and a 4-day forecast formed by multiple models (red) trained over the previous three months with $a_T \in \{1, 2, \ldots 96\}$. The relative MAE over the whole test period is 1.61% and the relative MAE over the peak periods is 1.12%. Generic BW units are used.

Several extensions and improvements of this forecast model are possible. The model can be extended to account for the dependencies between multiple tunnels, where the current value depends not only on the same TE tunnel's past values but also the past values of all other TE tunnels. Such a model would have a very large parameter space and would need additional penalization or model selection to work well in practice. Another generalization is an online version of this model, where the model is retrained and updated on every sample. We also train separate models at each forecast horizon, but these models are interlinked and it may be more efficient to train a joint model that accounts for dependencies between them. We did not pursue these extensions in this paper, since the model described here already gives a good fit and runs quickly enough for our purpose.

### E. Using Machine Learning and other Feedback to Optimize at Various Time Scales

SDN controller implementation brings along real-time network data and the capability of data-driven analytics for proactive closed-loop network management. Further, machine learning techniques can be applied on different time scales depending on the network states and scenarios.

**Sub-Second Time Scale:** Here we do not do any real-time computations but rely on a Fast Reroute (FRR) mechanism to temporarily bypass a failed path to a pre-computed backup path [19] within an order of tens of milliseconds of detecting a failure. In current practice we typically use a static back-up tunnel or FRR bypass path for every link bundle in the network. It is important to select a shortest possible FRR bypass path that has enough capacity under all traffic and failure conditions. With machine learning (ML) based timely traffic prediction we can periodically re-adjust and optimize these FRR bypass paths. It is to be noted that ML computation and associated optimized FRR path computations are only done in seconds-to-minutes time scale, but such computations improve performance in the sub-second time scale.

**Seconds-to-minutes Time Scale:** We have demonstrated recently [8] that an SDN controller can retrieve real-time network data, make and execute optimal layer 3 TE tunnel changes over fixed IP links in a sub-minute interval. Real-time machine learning can make the changes proactively and is therefore less disruptive to customers.

**Minutes-to-hours Time Scale:** In addition to being able to reroute TE tunnels over fixed IP links, we can also use the flexibility of CD ROADMs and DFCC to create new IP links, delete or reroute an existing IP link based on changing network and traffic conditions. Machine learning predictions allow us to do this in a proactive manner rather than in a reactive manner based on real-time feedback.

**Days, weeks and months Time Scale:** The introduction of a multi-layer SDN controller fundamentally changes the network capacity augmentation process from relatively disjointed L0 and L3 capacity planning to an integrated multi-layer planning; and from a single long-term planning horizon to include a much shorter planning timescale. Here the main need is to ascertain how much extra resources are needed within days in addition to weeks and months of ordering interval. The resources include router ports, optical transponders and Regenerators. We need to simulate many failure scenarios in an integrated multi-layer fashion based on future traffic predictions with a timescale of days and weeks. Machine learning plays a critical role here by accurately predicting the entire Traffic Matrix in time scales of days and weeks.

**Optimization Methodology:** We have developed efficient heuristics to optimize over many different failure scenarios and joint global optimization of optical and IP layers using the flexibility of CD ROADMs and DFCC [6]. The heuristics provide a close to optimal solution and reduces the execution times for a large network from 10s of minutes to a few seconds in the seconds-to-minutes and minutes-to-hours time scales. Also while working on the capacity planning for the days-to-months time scale, the heuristics reduce the execution time from several hours (or even days) to a few minutes. We provide a short overview here.

**Seconds-to-minutes time scale:** In this regime, we cannot change the IP links, their capacity or their routing over the optical layer. In this setting, the most efficient routing of TE tunnels can be achieved by multi-commodity flow [20] which requires arbitrary splitting of TE tunnels. Since we have a practical constraint on how TE tunnels can be split and different latency constraints for different traffic types, an alternative method is to use Constrained Shortest Path (CSPF) routing. By careful orchestration of the order of tunnel routing, we achieve almost the same efficiency as multi-commodity flow but it runs much faster.

**Minutes-to-hours time scale:** In this regime, we can create new IP links by connecting Tails and Regens, create new Tails by combining a Router port and a transponder using DFCC, or alternatively disconnect existing links or existing Tails and make their components free for future use. At any given instant, we keep the minimum number of IP links connected to be able to carry traffic while honoring the latency and loss constraints of each class and keep as many resources (Tails and regens) free as possible for future use. If there is a traffic surge event or failure event that no longer allows traffic to be carried appropriately, then one or more IP links are created by connecting free Tails (and Regens if needed). Many possible IP link choices are evaluated and the one that leaves the maximum amount of spare resources for future use is selected. Alternatively, we periodically check if there are too many IP links in the network and if appropriate, remove one or more IP links with the objective of maximizing spare resources for future use while satisfying latency requirements.

**Days, weeks and months timescale:** We consider a specific time period $T_1$ in the future at which new resources may be added and another time period $T_2$ up to which we have to live with those resources. We simulate many failure scenarios (Router failures, fiber cuts and optical equipment failures) and traffic surge scenarios (based on past observations and machine learning based predictions up to time $T_2$) and make sure that there are enough resources in the network (Tails and Regens) to satisfy the latency and loss constraints of each traffic class for each scenario. If for any scenario, the existing resources (plus resources added as a result of satisfying previous scenarios) are not enough then the minimal amount of additional resource needs are added. The whole process is repeated with a different order of scenario consideration. At the end of the process, the total amount of minimal additional resource need is identified. This amount of resource is added at time $T_1$ and the network needs to live with it up to time $T_2$.

### F. Comparison of traditional design and operation of IP/Optical Networks with that based on Machine Learning and Joint Multilayer Optimization

**Improved Efficiency:** Table I shows improved efficiency and cost reduction with machine learning based traffic prediction combined with joint multilayer optimization at a large ISP network with a large number of MPLS-TE tunnels (in the range of 4000 to 6000). All numbers shown are generic normalized values and are only to be used to compare among the different scenarios. Analysis is at the one month time scale to determine the minimal number of resources needed to satisfy all potential failure and traffic loss scenarios over that period of time. For the purpose of this analysis we assume that we have to optimize over two types of resources, Tails and Regens. The cost is given in units of 100 GE

(Gigabit Ethernet) Tails and for the purpose of illustration, it is assumed that the cost of a 100 GE Regen is 40% that of a 100 GE Tail. We consider four cases:

1. No Machine Learning, IP Layer Optimization Only, and fixed IP to Optical mapping: Requires extra capacity to account for traffic uncertainties.

2. Addition of Machine Learning: In this scenario, we have more precise knowledge of time-of-day and day-of-week traffic variation allowing for a tighter network design.

3. Addition of Joint Multilayer Optimization with a Fixed IP Layer Topology: We use the same set of IP links under all conditions as in the above two scenarios but the capacity of an IP link can be readjusted (e.g., using 3 100GE wavelengths vs 2 100GE wavelengths for the same IP link). Under a given failure scenario if a subset of components fail then the remaining non-failed components can be reused to enhance the capacity of an existing IP link. As one example, if there is an IP link from A to B and there is a fiber cut on the path but Tails at the endpoints stay intact then the Tails may be re-connected over a longer path on the fiber network (possibly requiring a free Regen on that longer path) to recreate the same IP link capacity. Alternatively, if there are other IP links between A and C and between B and D and there are free Tails at C and D then the capacity of the A-C and B-D IP links can be enhanced by re-using the Tails at A and B.

4. Addition of Dynamically Changing IP Layer Topology as traffic changes: Here for each failure scenario and traffic surge scenario, we rearrange the number of IP links and their routing over the optical network in order to optimally use the Tail and Regen resources. Furthermore, we typically have two Routers in every office and if one fails then we can use DFCC to create a new Tail using a port of the other Router and re-using the transponder.

Table I. Normalized View of Efficiency Gains with Machine Learning for Traffic Prediction Combined with Multi-Layer Optimization

| Scenario | # of 100 GE Tails | # of 100 GE Regens | Cost |
|---|---|---|---|
| 1. No Machine Learning, IP Layer Optimization Only, fixed IP to Optical mapping | 1,000 | 100 | 1,040 |
| 2. Addition of Machine Learning at long time-scale | 910 | 90 | 946 (-9%) |
| 3. Addition of Joint Multilayer Optimization with a Fixed IP Layer Topology | 810 | 80 | 842 (-19%) |
| 4. Addition of Dynamically Changing IP Layer Topology as traffic changes | 640 | 110 | 684 (-34%) |

We see that if we just do Machine Learning (ML) providing more accurate traffic prediction, we have about 9% saving. Next if we combine ML with joint multi-layer optimization where the mapping between IP links and optical resources may be readjusted and IP link capacity may be changed but with the same set of IP links (i.e., fixed IP layer topology), we get about 19% saving. Finally if we combine ML with full multi-layer global optimization (mapping of IP links to optical resources can change and IP links themselves can change as network condition changes), we get about 34% saving. It is interesting to see that the last case typically uses many more IP links (but significantly lower capacity per IP link) and can greatly reduce the number of Tails. However, since it often uses longer distance IP links, it needs to use more regens (but the overall cost of Tails + regens is lower).

**Less disruption to customer traffic with proactive Machine Learning Based Approach:** We considered a tight and highly optimized network design of a large ISP network, simulated failures and traffic surges and made the following observations:

- If the same static Fast Reroute (FRR) Backup path is used irrespective of traffic changes, then some traffic losses may occur when a failure happens near the peak traffic period as the pre-defined FRR backup path may not have sufficient capacity. It was possible to avoid these losses by proactively changing the FRR backup paths based on traffic changes predicted by machine learning. Alternatively, if our goal is to always avoid traffic loss then the static FRR paths would be more expensive compared to dynamically changing FRR paths based on future traffic changes predicted by ML.

- With a ROADM network controller, one can add and/or re-arrange a wavelength much faster than traditional manual and static methods but it still takes about 2-3 minutes to complete. Therefore, if one tries to change the IP layer topology during a peak-traffic period based on reactive real-time-based traffic observation, one will experience some traffic loss. Using machine learning prediction, this traffic loss could be avoided by making the IP layer topology changes about 20 minutes before the traffic surge. Again, if we want to avoid traffic loss but with a static reactive method then we would need more resources, thereby increasing cost.

**Ability to offer more efficient Bandwidth Calendaring Service:** Due to temporal variation and asymmetry of traffic matrix, there is usually significant amount of spare capacity left in the network that can be used for offering a flexible service with temporary capacity needs such as a bandwidth calendaring service [3, 5]. As this type of service may be offered in a matter of hours and days that is much shorter than any capacity planning cycle, the knowledge of near-term traffic pattern through machine learning can significantly improve the feasibility and efficiency of offering such service.

## III. USING MACHINE LEARNING TO PREDICT OPEN ROADM OPTICAL PATH PERFORMANCE

### A. Problem Description

Figure 8 shows a typical wavelength over a ROADM network. Each rectangle represents a separate building and no building in the figure has more than one ROADM.

ROADMs support Layer 1 services, such as private lines, and provide transport for higher layer services. Each ROADM is connected to one or more other ROADMs with one or more pairs of fibers. A Layer 1 wavelength can be set up between two transponders. Each transponder is connected to a nearby ROADM, and the wavelength is then routed through the ROADM network. Transponders are distinct from both a ROADM and an IP Router.

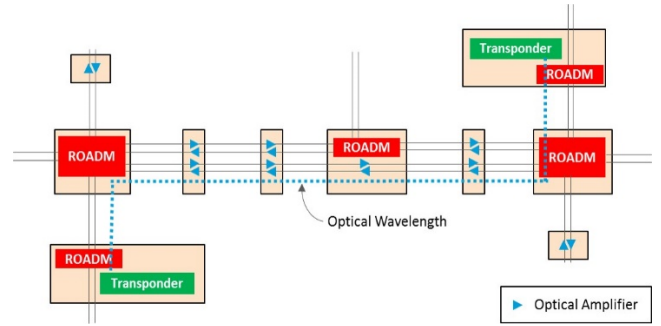

Fig. 8. Portion of a Sample ROADM Network.

Various factors can affect the quality of the optical signal. Imperfections in the fiber can add noise, and there will be signal distortions as the signal passes through equipment such as ROADMs and amplifiers and over distance. Thus, it is important to verify that a new wavelength will meet performance standards before putting it into service. Because Open ROADM networks can include equipment from multiple vendors, we cannot use a proprietary single-vendor tool to analyze new wavelength paths. Instead, we propose a machine learning model to predict optical performance.

### B. Model Features

In order to construct a machine learning model, we compile all available data for every optical wavelength in an existing ROADM network. We then distill this data into a set of 26 input features for each wavelength, or data sample. These features include data rate, fiber type, frequency, length of path, margin, measured fiber loss, measurement date, number of amplifiers in the path, number of pass-through ROADMs, ORL (Optical Return Loss), end-of-path OSNR (Optical Signal to Noise Ratio), and PMD (Polarization Mode Dispersion). We estimate the OSNR of each fiber section based on launch power, amplifier noise and measured span loss. We then combine these fiber section estimates to estimate the end-to-end path OSNR. In cases where regeneration is needed, we treat the sections between regeneration points as separate wavelengths.

As a measure of service quality, we wish to predict the Pre-FEC (forward error correction) Bit Error Rate for each wavelength in each direction. Since the BER values span several orders of magnitude, we use $\log_{10}(\text{BER})$ as the quantity to be estimated.

### C. Machine Learning Analysis

We apply a variety of machine learning algorithms and compare their performance, but the dataset (approximately 2700 samples) is smaller than typical machine learning applications, and the quality and quantity of the data turns out to be more of a limiting factor than the statistical methods we use. We focus on penalized and ensemble regressions, which are well suited for small-scale data such as this. We do not consider more sophisticated models like deep neural networks, as they are likely to overfit due to the small sample size and diverse types of features in the data. We use scikit-learn [18], a free open-source Python library that contains industry-standard implementations of these and many other machine learning models.

The machine learning methods we consider broadly fall into three categories: penalized linear regressions, nonlinear regressions and ensembles of regression trees. We denote the set of features by $X$ and the output values $(\log_{10}(BER))$ by $y$.

In the first category, we consider ridge regression and LASSO. These are traditional models that are easy to interpret and serve as a baseline for the other methods. They both estimate coefficients $\beta$ in a model of the type $\min_{\beta} \|y - \beta X\|_2^2 + a\|\beta\|_p^p$, with $p = 2$ (ridge) or $p = 1$ (LASSO). The latter encourages sparsity in $\beta$, as would be expected if most features have no impact on the BER, while the former reduces instability in estimating $\beta$ when the features are highly correlated.

Among nonlinear regression models, we look at the performance of quadratic LASSO, Gaussian process regression and the multilayer perceptron. Quadratic LASSO is a simple variant of LASSO using features of the form $x_i x_j$ for every pair $(i, j)$, giving a total of 676 features. Gaussian progress regression was described in Section II.D, while the multilayer perceptron is a classical, fully-connected neural network.

Among ensemble models, we apply gradient boosted regression trees and random forests. A regression tree is a piecewise linear regression that iteratively splits the data according to an error criterion and fits separate regressions to each portion of the split data. Both ensemble methods train several different regression trees over different subsets of the features, and take an average over all of the trees to obtain a final estimate.

More details of all of these algorithms can be found in [16].

### D. Model Performance

For each machine learning model, we consider 50 random splits of the data, each with 2/3 of the data used for training the model and 1/3 used for testing the model. Various hyper-parameters of each model (e.g., the penalty factor in ridge and LASSO regressions) are optimized by choosing random values on each split and taking the best one. To measure the performance of an algorithm, we compute the mean squared error (MSE) across all points, the MSE for only the points with high BERs above $10^{-6}$ (which we denote HMSE), and the MSE across the points with the 10% worst errors (denoted WMSE), averaged over all 50 splits of the dataset. In practice, we want the model to have good ballpark estimates of BER (not necessarily very precise ones) and are especially interested in the measurements with higher BERs, so the HMSE is more important than the MSE.

Table II shows the performance of the different machine learning models according to these error criteria. The units on the error rates are in terms of the log(BER). For example, an MSE of 1.06 means that on average, the predicted BER and the actual BER differ by one order of magnitude. Based on the MSE criteria, the random forest model and the gradient boosted models perform the best. Based on the HMSE criteria, the random forest model and the Gaussian process regression models perform the best. The random forest model consistently achieves the best overall error rates based on both criteria, at the cost of a higher model complexity and less interpretability. The random forest takes the average of several regression trees over randomly chosen subsets of the features, any of which may contribute to the model reaching a particular predicted BER value. Usually BER estimates within one or two orders of magnitude of the actual BER are good enough to classify wavelength performance as good or bad and the random forest model meets this criteria. We also found that standardizing each feature before applying the model, a common data preprocessing technique, does not improve the performance due to the diverse mix of continuous and discrete features.

Table II. Performance of Different Machine Learning Models

| Model | MSE | HMSE | WMSE |
|---|---|---|---|
| Ridge regression | 1.06 | 3.32 | 5.80 |
| LASSO regression | 1.15 | 3.63 | 6.25 |
| LASSO with quadratic features | 0.83 | 2.30 | 5.19 |
| Multilayer perceptron | 0.94 | 2.91 | 6.12 |
| Gaussian process regression | 0.90 | 1.87 | 5.90 |
| Gradient boosted regression trees | 0.81 | 2.08 | 5.18 |
| Random forest regression trees | 0.81 | 1.86 | 5.14 |

For the random forest, the predicted and actual BER across one of the training/testing splits is shown in Figure 9 where each tick mark represents one order of magnitude.

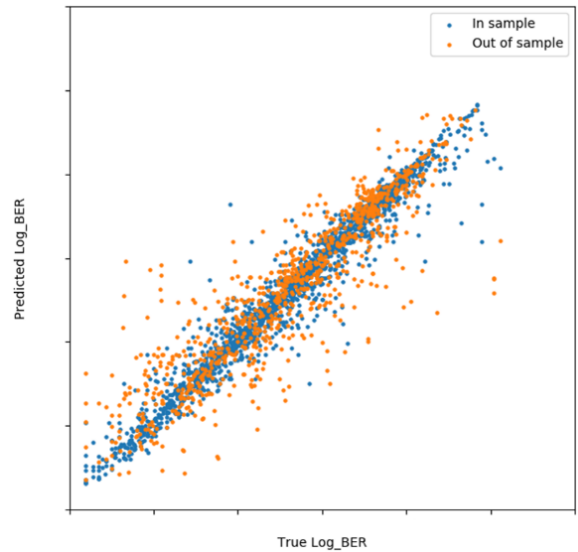

Fig. 9. Machine Learning Model Performance (Random Forest).

### E. Importance of Features

In a random forest model, the importance of a given feature can be measured by randomly permuting values of the feature and measuring how much the regression error increases. This is used to form a score for each feature known as the Gini importance (see [16] for details). We apply this methodology here, with the importance scores averaged over all 50 models and splits of the data and normalized on a scale from 0 to 100, with 100 indicating the most important feature.

In Figure 10, most of the contribution to the model accuracy comes from the top four features, which are the data rate (40G or 100G), the path length, the OSNR and the wavelength frequency respectively. It is a little surprising to observe that OSNR has only the third highest impact even though it is directly correlated to path length. It appears that in addition to depending on path length, it also depends on several other factors that are less important and thereby the impact of OSNR on BER is diluted. The remaining 22 features have little effect on the BER. We train the random forest with only the top 10 features (importance score over 2.00) and obtain MSEs of 0.86/1.87/5.40, which are not far off from the errors in Section IIID.

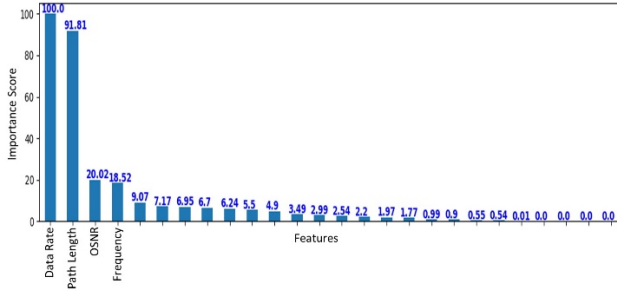

Fig. 10. Illustration of Relative Importance of Machine Learning Features.

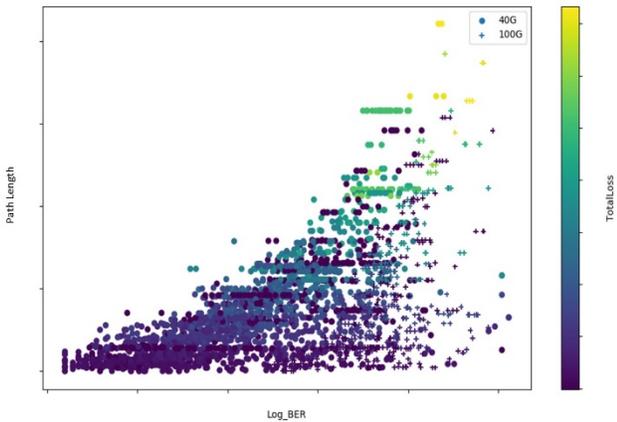

Fig. 11. Effect of Data Rate, Path Length and Loss on BER.

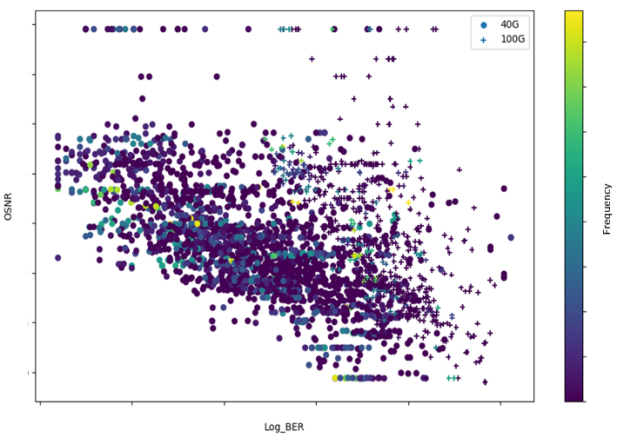

Fig. 12. Effect of Data Rate, OSNR and Wavelength Frequency on BER.

Figure 11 shows the BER as a function of the data rate, path length and total loss. Figure 12 shows the BER as a function of the data rate, OSNR, and wavelength frequency. These plots demonstrate the complex, nonlinear structure of the data as well as the fact that no single feature by itself is sufficient for predicting the BER. It also indicates why the random forest is able to outperform a simple linear regression model.

### F. Application

The model can be incorporated as a feature of the Path Compute Engine (PCE) within the SDN Controller to verify that all new Open ROADM wavelengths meet optical performance standards. The PCE can generate a proposed path for each new wavelength request, and then invoke the Machine Learning model to predict the optical performance of the proposed path. If the path meets optical performance standards, then the wavelength can be deployed into the Open ROADM network. Otherwise, the PCE can generate an alternative path and try again. Periodically, the SDN Controller can use the latest network data to retrain the Machine Learning model and then update the model parameters.

### G. Observations

The proof of concept study demonstrates that it is possible to create a machine learning model to predict the optical performance of ROADM wavelengths, specifically pre-FEC bit error rate, with reasonably good accuracy without knowing many of the details of the optical line or fiber. In particular, the model is able to do this with fewer features and far less data than typical machine learning applications.

The next step will be to extend the machine learning model to predict optical performance of wavelengths in the new Open ROADM network. This model can be implemented as a microservice as part of the Path Compute Engine within the SDN Controller. The Open ROADM version of the model may have slightly different features due to differences in the new network, but the general approach should be similar. While the data supporting this study comes from a single vendor network, other vendor equipment should provide similar data with similar interpretation and thus the methodology can be applied in a multi-vendor environment. In addition, as Open ROADMs are model-driven, performance data from other optical plug-ins can be included, if needed, to further enhance the model and predictability.

In addition to planning paths for new wavelengths, the SDN Controller can also use the Machine Learning model to monitor the optical performance of existing wavelengths and move them to better paths as conditions evolve.

## IV. SUMMARY

We have described two applications of machine learning for managing IP and Optical networks. The first application allows significant cost saving by combining machine-learning-based long-term traffic prediction with global optimization of IP/Optical layers using CD ROADMs and DFCC devices. The second application uses machine-learning-based short-term traffic prediction to allow proactive network changes to reduce customer traffic disruptions and opens an opportunity to offer flexible services based on dynamic capacity needs. The second application enables the selection of improved ROADM paths based on the latest optical performance data. Both applications can be efficiently implemented using an SDN controller.

These methodologies can be extended to different network settings depending on technology evolution, network data availability, and maturity of machine learning.

## ACKNOWLEDGMENT

The authors would like to thank their colleagues, Martin Birk and Kathy Tse, for providing their expert knowledge in Open ROADM and Optical Transport Network as well as suggesting the optical path performance use case. They would also like to thank Kathy Meier-Hellstern for carefully reviewing the paper and providing many valuable comments. Finally, thanks are due to the anonymous reviewers and the Editor-in-Chief for their excellent comments and suggestions that significantly improved the quality of the paper.

## REFERENCES

[1] Z. Fadlullah, F. Tang, B. Mao, N. Kato, O. Akashi, T. Inoue, and K. Mizutani, "State-of-the-Art Deep Learning: Evolving Machine Intelligence Toward Tomorrow's Intelligent Network Traffic Control Systems," IEEE Commun. Surveys & Tutorials, Vol. 19, Issue 4, 2017, pp. 2432-2455.
[2] F. Musumeci, C. Rottondi, A. Nag, I. Makaluso, D. Zibar, M. Ruffini and M. Tornatore, "A Survey on Application of Machine Learning Techniques in Optical Networks," ar.XIV:1803.0797vl [cs.NI], March 2018, pp. 1-20.
[3] J. Donovan and K. Prabhu, "Building the Network of the Future", CRC Press, First Edition, ISBN 1138631523, September 2017.
[4] M. Birk, G. Choudhury, B. Cortez, A. Goddard, N. Padi, A. Raghuram, K. Tse, S. Tse, A. Wallace and K. Xi, "Evolving to an SDN-Enabled ISP Backbone: Key Technologies and Applications", in IEEE Communications Magazine, Vol. 54, Issue 10, October 2016, pp. 129-135.
[5] L. Gkatzikis, S. Paris, I. Steiakogiannakis, and S. Chouvardas, "Bandwidth calendaring: Dynamic services scheduling over Software Defined Networks," IEEE ICC 2016.
[6] G. Choudhury, M. Birk, B. Cortez, A. Goddard, N. Padi, K. Meier-Hellstern, J. Paggi, A. Raghuram, K. Tse, S. Tse and A. Wallace, "Software Defined Networks to greatly improve the efficiency and flexibility of Packet IP and Optical Networks", ICNC 2017, January 26-29, 2017, Silicon Valley, CA, USA.
[7] Y. Li, G. Shen and L. Peng, "Impact of ROADM colorless, directionless, and contentionless (CDC) features on optical network performance," Invited paper, IEEE/OSA Journal of Optical Communications and Networking (Volume: 4, Issue: 11, Nov. 2012, pp. B58-B67).
[8] S. Tse and G. Choudhury, "Real-Time Traffic Management in AT&T's SDN-Enabled Core IP/Optical Network," Invited presentation at OFC 2018, March 11-15, San Diego, CA, USA.
[9] Open ROADM MSA, http://openroadm.org
[10] TransportPCE,https://wiki.opendaylight.org/view/TransportPCE:Main
[11] M. Birk, "AT&T's Direction Towards a Whitebox ROADM", Open Network Summit 2015, Session: SDN for Service Provider Networks: Transport SDN, June 17th 2015, Santa Clara, CA.
[12] L. Barletta, A. Giusti, C. Rottondi and M. Tornatore, "QoT Estimation for Unestablished Lightpaths Using Machine Learning," OFC 2017, March 2017, Los Angeles, CA, USA
[13] G. Zervas, M. Leenheer, L. Sadeghioon, D. Klonidis, Y. Qin, R. Nejabati, D. Simeonidou, C. Develder, B. Dhoedt and P. Demeester, "Multi-Granular Optical Cross-Connect: Design, Analysis and Demonstration," IEEE/OSA Journal of Optical Communications and Networking, Vol. 1, Issue 1, June 2009.
[14] I. Kim, P. Palacharla, X. Wang, D. Bihon, M. Feuer and S. Woodward, "Performance of Colorless, Non-directional ROADMs with Modular Client-side Fiber Cross-connects," OFC 2012, Los Angeles, CA, USA.
[15] C. E. Rasmussen and C. K. I. Williams, "Gaussian Processes for Machine Learning," MIT Press, Cambridge, MA, 2005.
[16] T. Hastie, R. Tibshirani and J. Friedman, "The Elements of Statistical Learning: Data Mining, Inference, and Prediction, Second Edition", Springer, New York, NY, 2009
[17] J. D. Hamilton, "Time Series Analysis", Princeton University Press, Princeton, NJ, 1994
[18] F. Pedregosa, G. Varoquaux, A. Gramfort, V. Michel, B. Thirion, O. Grisel, M. Blondel, P. Prettenhofer, R. Weiss, V. Dubourg, J. Vanderplas, A. Passos, D. Cournapeau, M. Brucher, M. Perrot and É. Duchesnay, "Scikit-learn: Machine Learning in Python," The Journal of Machine Learning Research, 12:2825–2830, 2011
[19] P. Pan, G. Swallow and A. Atlas, "Fast Reroute Extensions to RSVP-TE for LSP Tunnels," IETF RFC 4090, May, 2005.
[20] T. Cormen, C. Leiserson, R. Rivest, and C. Stein, "Introduction to Algorithms (3rd ed.)". MIT Press and McGraw–Hill, 2009.

## BIOGRAPHIES

**Gagan Choudhury** (gchoudhury@att.com) is a lead inventive scientist at AT&T Labs, Middletown, New Jersey. He received a Ph.D. in electrical engineering from the State University of New York at Stony Brook in 1982. His research interests are in the optimization, analysis, and design of software defined networks and mobility networks. He is an IEEE Fellow (2009) and became an AT&T Fellow in 2009 for "outstanding contributions to performance analysis and robust design and their application to improving the performance, reliability and scalability of AT&T's networks."

**David Lynch** (dflynch@att.com) is a lead inventive scientist at AT&T Labs, Middletown, New Jersey. He received an A.B. in mathematics and computer science from Susquehanna University and an M.S. and Ph.D. in operations research from Cornell University. Since joining AT&T in 1984, he has worked on a variety of network optimization applications, including private and public, voice and data, WAN and metro, and Layers 1 through 3. His research interests include development and implementation of telecommunications network optimization algorithms.

**Gaurav Thakur** (gt6510@att.com) is a principal inventive scientist at AT&T Labs, Middletown, New Jersey. He received a B.S. in mathematics from the University of Maryland in 2007 and a Ph.D. in applied mathematics from Princeton University in 2011. His research interests are in statistical signal processing, machine learning and harmonic analysis, as well as applications of these techniques to other fields.

**Simon Tse** (stse@att.com) is a director of inventive science at AT&T Labs, Middletown, New Jersey. He received a B.S. in engineering from Brown University, an M.S. and a Ph.D. from Harvard University in applied sciences, and an M.B.A. from the Wharton School of the University of Pennsylvania. He began his career in 1985 with AT&T Bell Laboratories. He currently manages a group of technical professionals in network topology designs, network traffic management, and software defined network controllers for multi-layer network resource and routing optimization.